\documentclass[Journal,10pt]{IEEEtran}

\usepackage{color}
\usepackage{float}
\usepackage{tabularx,colortbl}
\usepackage{setspace}
\usepackage{cite,graphicx,amsmath,psfrag,bm}
\usepackage{subfigure}
\usepackage[usenames,dvipsnames]{xcolor}
\usepackage{mathabx}
\usepackage{cancel}
\usepackage{epstopdf}
\usepackage{amsmath}
\usepackage{cancel}
\usepackage{caption}

\usepackage[process=auto]{pstool}
\usepackage[normalem]{ulem}
\graphicspath{{Figs/}}

\newcommand{\figref}{Fig.~\ref}

\ifCLASSINFOpdf
\else
\fi

\begin{document}

\title{Spread-Spectrum Selective Camouflaging \\ based on Time-Modulated Metasurface}

\author{Xiaoyi~Wang,~\IEEEmembership{Student Member,~IEEE}
        and Christophe~Caloz,~\IEEEmembership{Fellow,~IEEE}
\thanks{
The authors are with the department
of electrical engineering, Polytechnique Montr\'{e}al,
Montr\'{e}al, Qu\'{e}bec, Canada. (E-mail: xiaoyi.wang@polymtl.ca).}
}

\maketitle
\begin{abstract}
This paper presents the concept of spread-spectrum selective camouflaging based on time-modulated metasurface. The spectrum spreading is realized by switching the metasurface between the reflective states of a PEC mirror and a PMC mirror, using an array of microstrip patches connected to the ground via diode switches, according to a periodic pseudo-random noise sequence. As the spectrum spreading induces a drastic reduction of the power spectral density of the signal, the level of the scattered wave falls below the noise floor of the interrogating radar, and the object covered by the metasurface is hence perfectly camouflaged to a foe radar. Moreover, the object can be detected by a friend radar possessing the spread-spectrum demodulation key corresponding to the metasurface modulation, and this detection is robust to interfering signals. The proposed system is analyzed theoretically, and demonstrated by both full-wave simulation and experimental results.
\end{abstract}


\section{Introduction}
Electromagnetic camouflaging refers to concealment technologies whereby objects are made undetectable~\cite{Book:Stevens_2011_AnimalCamouflage,Book:Lynch_RF_Stealth}. It is widely present in nature, for instance in butterflies with wings mimicking leaves, jellyfishes with quasi-transparent bodies, and chameleons that adapt their colors and patterns to the environment~\cite{Book:Stevens_2011_AnimalCamouflage}. It is also produced by humans, for instance  in hunting or military suits and in radar-stealth aircraft and warships~\cite{Book:Lynch_RF_Stealth}.

Camouflaging is generally realized by altering the spectrum or power density of the waves scattered by the object to conceal. Such alteration may be accomplished in different manners, including bio-inspired paintings with dazzling or counter-shading patterns~\cite{Jour:Rune_2018_VisualCamouflage}, absorbing material coatings~\cite{Jour:APL_Wang_2011_AbsorbingMaterial,Jour:AM_Xia_2013_AbsorbingMaterial,Jour:JMMM_Lima_2008_RadarAbsorbingMaterial}, stealth shaping~\cite{Jour:TM_Bongeson_2004_RCSOptimaization}, and spectral power redistribution~\cite{Jour:Tennant_2004_Screen,Jour:Tennant_1997_Screen}.

In past decade, metasurfaces, the two-dimensional counterparts of voluminal metamaterials, have spurred major interest in both the scientific and engineering communities owing to their attractive features of small form profile, low loss, easy fabrication and unprecedented flexibility in controlling the amplitude, phase and polarization of electromagnetic waves~\cite{Jour:Nanophotonics_2018_Achouri_Metasurface}. A great diversity of metasurface applications have been reported to date, including for instance polarization transformation~\cite{Jour:PRX_jiang_2014_ControllingPolarization}, wavefront manipulation~\cite{Jour:PJ_Genevet_2014_WavefrontControl,Jour:PRL_Asadchy_2015_FunctionalMetamirrors,Jour:LSA_Karimi_2014_OAM,Jour:PRApp_Pfeiffer_2014_BesselBeam}, holography~\cite{Jour:NN_Zheng_2015_MetasurfaceHolograms,Jour:NC_YE_2016_NonlinearMetasurfaceHolography}, nonreciprocity~\cite{Jour:Sounas_2012_NonreciprocalMetasurface,Jour:Caloz_2018_PRApp_Nonreciprocity}, optical force carving~\cite{Jour:Grbic_2015_PRB_tractorbeam,Jour:Achouri_2019_TAP_OpticalForce}, and analog computing~\cite{Jour:Science_Silva_2014_MathOperation}.

Given their multiple benefits, metasurfaces have naturally been considered for electromagnetic camouflaging, based on absorption~\cite{Jour:TAP_1973_Emerson_EM_absorbers,Jour:JPDAP_li_2014_CircAbsorber,Jour:PRB_Valagiannopoulos_2015_PerfectMirrorAndAborber,Jour:OL_Wen_2014_AbsorptionMetamaterial}, scattering redirection~\cite{Jour:EM_1985_Raj_Scattering,Jour:SR_2016_Chen_Redirection_metasurface}, and cloaking~\cite{Jour:PRB_2009_Alu_MantleCloak,Jour:TAP_Dehmollaian_2019_Cloaking,Jour:Alu_AWPL_2014_CarpetCloak}.
However, these technologies are typically limited by issues such as narrow bandwidth, large aperture~\cite{Jour:SR_2016_Chen_Redirection_metasurface}, camouflaging size limitation~\cite{Jour:Fleury_2015_invisibility}, and camouflaging indifferentiation (indistinct camouflaging to all observers).

These issues are largely related to the time-invariant nature of the corresponding systems and the related fundamental physical bounds. Revoking the time invariance contraint in time-modulated metasurfaces opens up the possibility to break these bounds and achieve new functionalities~\cite{Jour:ArXiv_Caloz_2019_SpacetimeMeta}. A few related applications have already been reported, such as serrodyne frequency translation~\cite{Jour:Arxiv_Wu_2019_frequency_translation}, simplified architecture communication~\cite{Jour:CC_Cui_2019_CommMeta,Jour:AMT_Cui_2019_CommMeta,Jour:Res_Cui_2019_PrograCoding,Jour:Cui_NSR_2018_Comm}, direction-of-arrival (DOA) estimation~\cite{Conf:Wang_2019_APS_DOA}, nonreciprocity~\cite{Jour:OME_2015_Shalaev_Nonreciprocity,Jour:PRX_2018_shadrivov_STM,Jour:PRB_2017_Ramaccia_Doppler_cloak}, and analog signal processing~\cite{Jour:TAP_2019_Nima_STM}, to mention a few.

In this context, we present here a time-modulated metasurface active camouflaging technology based spread spectrum, which was introduced in~\cite{Conf:Wang_2019_APS_Camouflaging}, whereby the spectrum of the incident wave is spread into noise below the noise floor of a radar interrogator, while providing the extra features of selective camouflaging and interference immunity.

\section{General Concept}\label{Sec:Concept}

\begin{figure*}[h!t]
    \centering
    \captionsetup{justification=centering}
    \includegraphics{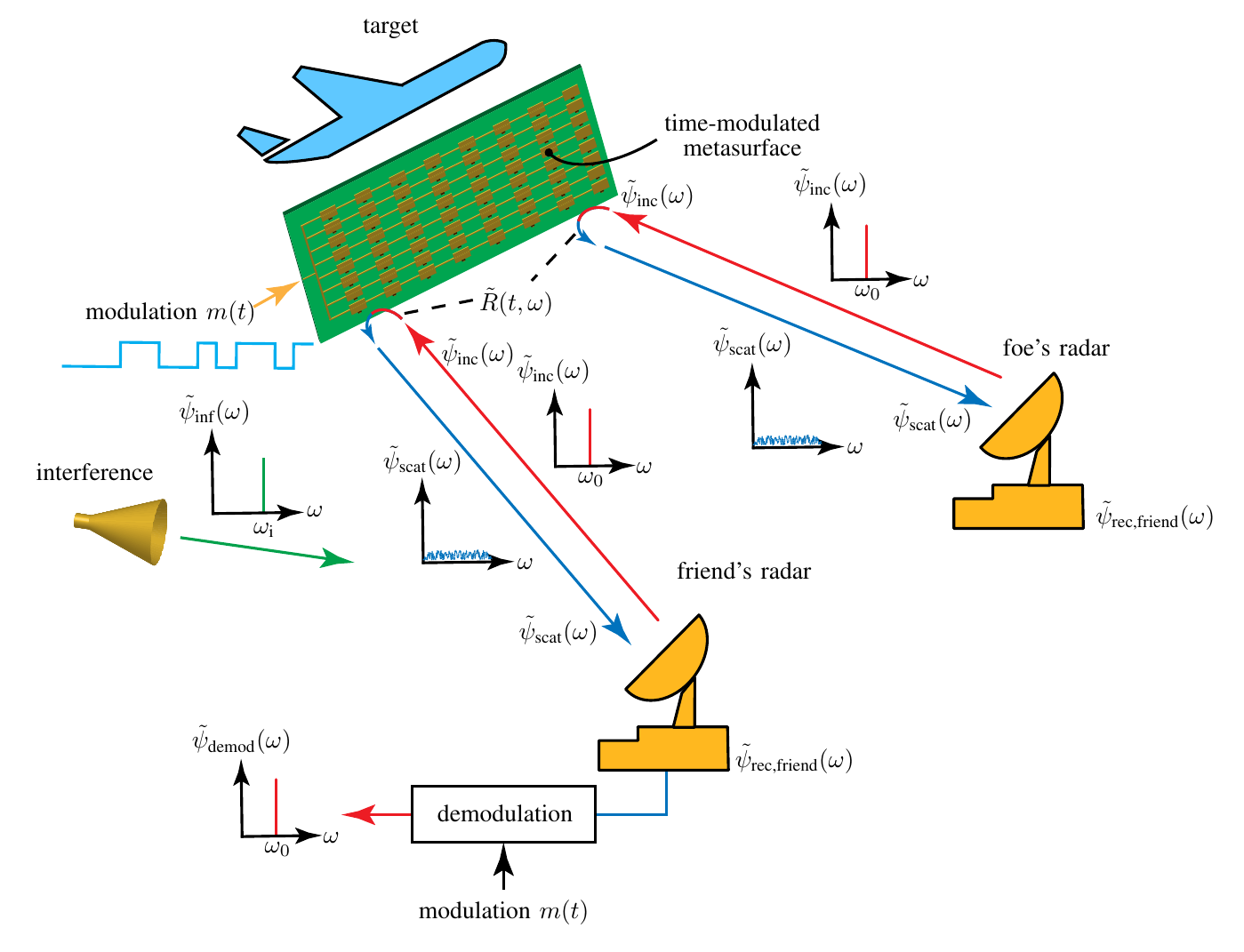}
     \caption{Proposed spread-spectrum time-modulated metasurface camouflaging.}
     \label{FIG:Schematic}
\end{figure*}

The proposed concept of spread-spectrum time-modulated metasurface camouflaging is illustrated in~\figref{FIG:Schematic}. The object to be detected is covered by a metasurface that is modulated by a temporal sequence $m(t)$, where $t$ is time, and that exhibits therefore the reflection coefficient $\tilde{R}(t,\omega)$, where $\omega$ is the angular frequency corresponding to the dispersive resonant nature of the scattering particles forming the metasurface. When a harmonic wave $\tilde{\psi}_\text{inc}(\omega)$ impinges on this structure, its spectrum gets spread out by the time variation into a noise-like signal, $\tilde{\psi}_{\text{scat}}(\omega)$, with extremely low power spectral density, so that the scattered wave is undetectable to any radar detector.

In addition to its basic camouflaging operation, the spread-spectrum time-modulated metasurface concept offers the smart functionality of selectivity, whereby the object can made detectable by fiends while being camouflaged to foes. The functionality is provided by leveraging the demodulation scheme of spread-spectrum used in wireless communications, with the spread-spectrum key corresponding to the time-varying reflection coefficient $\tilde{R}(t,\omega)$. Moreover, the spectrum spreading principle makes the friend detection highly robust to interference.

\section{Theory}\label{Sec:Theory}

\subsection{Spectrum Spreading}\label{sec:spectrum_spread}

\begin{figure}[h!]
	\centering
    \includegraphics{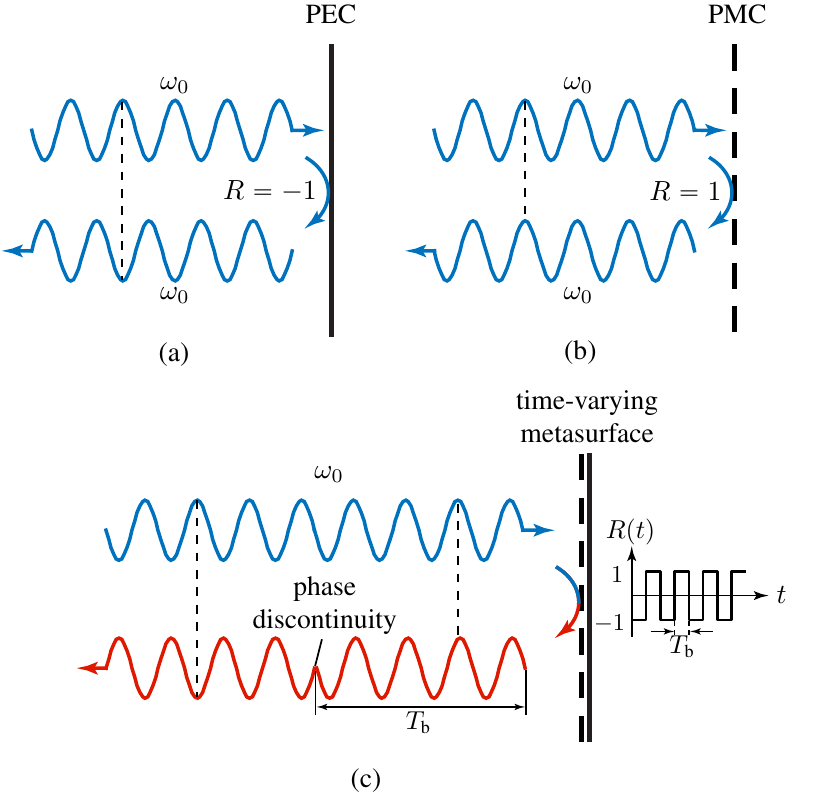}
	\caption{Principle of spectrum spreading by the time-modulated metasurface in \figref{FIG:Schematic}, assuming a time-harmonic interrogating wave of angular frequency $\omega_0$. (a)~Static PEC reflector, with reflection coefficient $R=-1$. (b)~Static PMC reflector, with reflection coefficient $R=1$. (c)~Time-varying metasurface reflector formed by repeatedly switching the reflection coefficient between the states (a) and (b), so as to make it dynamic, $R=R(t)$.}
	\label{FIG:SpreadSpectrum}
\end{figure}

The time-modulated metasurface spectrum spreading principle may be understood with the help of~\figref{FIG:SpreadSpectrum}, assuming a time-harmonic incident wave of angular frequency $\omega_0$. If the metasurface reduces to a static perfect electric conductor (PEC), as shown in \figref{FIG:SpreadSpectrum}(a), the incident wave is scattered back at $\omega_0$ after experiencing phase reversal on the reflector. Similarly, if the metasurface is a static perfect magnetic conductor (PMC), as shown in \figref{FIG:SpreadSpectrum}(b), the incident wave is scattered back at $\omega_0$, but without experiencing any phase alteration on the reflector. If the metasurface is now modulated so as to repeatedly switch between a PEC reflector and a PMC reflector, as shown in \figref{FIG:SpreadSpectrum}(c), it becomes dynamic, or time-varying, with reflection coefficient $R=R(t)$ varying between $-1$ and $+1$ at minimum time intervals $T_\text{b}$, where `b' stands for `bit'. The scattered waveform is still a time-harmonic wave of frequency $\omega_0$, but with phase reversal discontinuities that correspond to the switching between the PEC and PMC states.

The waveform of the scattered wave may be written as
\begin{equation}\label{EQ:timedomain}
\psi_\text{scat}(t)=\tilde{R}(t,\omega_0)\psi_\text{inc}(t)=\tilde{R}(t,\omega_0)e^{j\omega_0t},
\end{equation}
whose spectrum is
\begin{align}\label{EQ:frequencydomain}
\begin{split}
\tilde{\psi}_\text{scat}(\omega)&=\frac{1}{2\pi}\int_{-\infty} ^{+\infty} \tilde{R}(t,\omega_0)e^{j\omega_0t}e^{-j\omega t} dt \\
&=\frac{1}{2\pi}\int_{-\infty} ^{+\infty} \tilde{R}(t,\omega_0)e^{-j(\omega-\omega_0) t} dt \\
&=\tilde{\tilde{R}}(\omega-\omega_0,\omega_0),
\end{split}
\end{align}
where $\tilde{\tilde{R}}(\omega,\omega_0)$ is the Fourier transform of $\tilde{R}(t,\omega_0)$, and $\omega_0$ is to be considered as a constant parameter. The last result of~\eqref{EQ:frequencydomain} shows that the spectrum of the scattered wave is the spectrum of the modulated metasurface reflection coefficient, shifted to the frequency of the incident wave. Therefore, the spectrum of the incident wave is spread out into the spectrum of the modulation, with center frequency $\omega_0$ and power spectral density corresponding to $\tilde{R}(t,\omega_0)$, the spreading corresponding to the introduction of the aforementioned phase discontinuities. Camouflaging according to specification will then be realized by properly designing $\tilde{R}(t,\omega_0)$ in terms of the parameters of the modulating sequence $m(t)$.

\subsection{Selectivity}\label{sec:selectivity}

A friend radar, knowing the spread-spectrum key of the metasurface, $\tilde{R}(t,\omega_0)$, may demodulate the spread signal $\psi_\text{scat}(t)$ by the simple post-processing division
\begin{equation}\label{EQ:harmonicDemod}
\psi_\text{demod}(t)=\psi_\text{scat}(t)\frac{1}{\tilde{R}(t,\omega_0)}=e^{j\omega_0t},
\end{equation}
where the second equality follows from~\eqref{EQ:timedomain}. Thus the friend radar detects the object that is camouflaged to foe radars, which do not possess the metasurface spread-spectrum key. The proposed metasurface camouflaging technology is thus selective in the sense that it may be restricted to undesired interrogators.

\subsection{Interference Immunity}\label{sec:interf_imm}
The demodulation used for selectivity automatically brings about the extra useful feature of immunity to interference. In the presence of an interfering signal, $\psi_\text{int}(t)$, represented on the left in \figref{FIG:Schematic}, the signal detected by the foe radar is
\begin{align}
\begin{split}
\psi_\text{foe}(t)&=\psi_\text{scat}(t)+\psi_\text{int}(t)\\
&=\tilde{R}(t,\omega_0)\psi_\text{inc}(t)+\psi_\text{int}(t),
\end{split}
\end{align}
and the interference further alters the signal received by the foe radar.

In contrast, the signal detected by the friend radar, after its demodulation section, is
\begin{subequations}\label{EQ:interfdemoduT}
\begin{align}
\begin{split}
\psi_\text{friend}(t)&=\psi_\text{foe}(t)\frac{1}{\tilde{R}(t,\omega_0)}\\
&=\left(\tilde{R}(t,\omega_0)\psi_\text{inc}(t)+\psi_\text{int}(t)\right)\frac{1}{\tilde{R}(t,\omega_0)}\\
&=\psi_\text{inc}(t)+\frac{\psi_\text{int}(t)}{\tilde{R}(t,\omega_0)} \\
&=\psi_\text{inc}(t)+\psi_\text{int}(t)\tilde{Y}(t,\omega_0),
\end{split}
\end{align}
with
\begin{equation}
\tilde{Y}(t,\omega_0)=\frac{1}{\tilde{R}(t,\omega_0)},
\end{equation}
\end{subequations}
where the effect camouflaging is removed, as in Sec.~\ref{sec:selectivity}, and the interfering wave is multiplied by the inverse of the reflection coefficient. Assuming that $\tilde{R}(t,\omega_0)$ oscillates between $-1$ and $+1$, as mentioned in Sec.~\ref{sec:spectrum_spread}, so does $\tilde{Y}(t,\omega_0)$, and the two functions are exactly the same, i.e., $\tilde{Y}(t,\omega_0)=\tilde{R}(t,\omega_0)$. As a result, the spectrum of the signal detected by the friend radar reads
\begin{align}\label{EQ:interfdemoduF}
\begin{split}
\tilde{\psi}_\text{friend}(\omega)
&=\tilde{\psi}_\text{inc}(\omega)+\tilde{\tilde{Y}}(\omega,\omega_0)*\tilde{\psi}_\text{int}(\omega) \\
&=\tilde{\psi}_\text{inc}(\omega)+\tilde{\tilde{R}}(\omega,\omega_0)*\tilde{\psi}_\text{int}(\omega).
\end{split}
\end{align}
This results shows that if the bandwidth of the interfering signal is smaller than that of the modulation, as is most common in practice, then that signal is spread out and gets ``camouflaged'' to the friend, and thence practically harmless to it.

In practice, as will be seen in the experimental part, the magnitude of the reflection coefficient is slightly less than~$1$ due to dissipative loss, i.e., $|\tilde{R}(t,\omega_0)|$ is slightly smaller than~$1$, and therefore, $|\tilde{Y}(t,\omega_0)|$ is slightly larger than~1, which tends to increase the effect of the interference. So, there is an antagonism between the reduction of the interference effect from the demodulation process and the increase of the interference effect due to the issue just mentioned. A good design, with $|\tilde{R}(t,\omega_0)|$ close to~$1$ will ensure that the former effect largely dominates the latter.

\subsection{Validity Condition of the Reflection Coefficient Description}

Particular attention must be paid to the precise meaning of the function $\tilde{R}(t,\omega_0)$. This expression seems \emph{a priori} absurd since it
is meant to represent a \emph{time-varying transfer function}, whereas the concept of transfer function is fundamentally restricted to linear \emph{time-invariant} systems~\cite{Jour:Lathi_2005_LinearSysSignals}. However, the expression $\tilde{R}(t,\omega_0)$ \emph{does} make perfect sense under the condition that the metasurface modulation occurs on a time scale, $T_\text{b}$ [\figref{FIG:SpreadSpectrum}(c)], that is much larger than the dispersion or memory time scale, $T_\text{d}$, which is naturally itself larger  than the interrogating signal period, $T_0=\omega_0/(2\pi)$, i.e.,
\begin{equation}\label{EQ:condition}
T_\text{b}\gg T_\text{d}>T_0.
\end{equation}

Under this condition, which assumes harmonic excitation ($T_0$) and discrete reflection switching ($T_\text{b}$), the system may indeed be \emph{considered} purely dispersive (without time variance) on the time scale $t<T_\text{b}$, and purely time-variant (without dispersion) on the time scale $t>T_\text{b}$, as implicitly considered in Sec.~\ref{sec:spectrum_spread} by considering $\omega_0$ as a constant parameter. Let us explain this in some detail.

Its reflection coefficient may then be written in terms of the purely linear time-invariant dispersive transfer function $\tilde{R}(\omega)$, where the mention of time variance has been accordingly suppressed, and we have
\begin{equation}\label{EQ:Rw_rel}
\tilde{\psi}_\text{scat}(\omega)=\tilde{R}(\omega)\tilde{\psi}_\text{inc}(\omega),
\end{equation}
corresponding in the time domain to\footnote{Note that the function $r(t)$ has the unit of inverse time (1/s), as required by the differential $d\tau$ in the convolution integral. This is in contrast to $\tilde{R}(\omega)$, which is, according to~\eqref{EQ:Rw_rel}, unitless.}
\begin{equation}
\psi_\text{scat}(t)=r(t)\ast\psi_\text{inc}(t)
=\int_{-\infty}^{t}r(t-\tau)\psi_\text{inc}(\tau)\,d\tau,
\end{equation}
where the upper integration limit $t$ ensures causality. Substituting $t-\tau\rightarrow\tau'$, and subsequently replacing the dummy variable $\tau'$ by $\tau$, yields
\begin{equation}
\psi_\text{scat}(t)
=\int_{0}^{\infty}r(\tau)\psi_\text{inc}(t-\tau)\,d\tau.
\end{equation}
Here, the lower integration limit corresponds to the onset of the system, while the upper integration limit corresponds to the duration of the impulse function $r(t)$, and hence to the transient time of the system. In practice, the function $r(t)$ may be truncated at a time $T_\text{d}$ where its average value has decayed to a negligible fraction of the maximum of $r(t)$, and the upper integration limit transforms then as $\infty\rightarrow{T}_\text{m}$.

\begin{figure}[h!]
    \centering
    \includegraphics{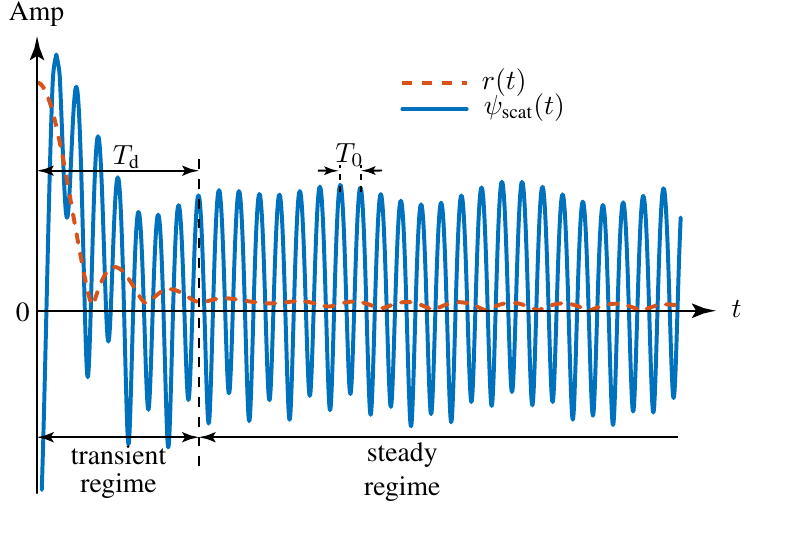}
     \caption{Transient regime and steady-state regime corresponding to the time-invariant dispersive nature of the metasurface within the time $T_\text{b}$ under the excitation $\psi_\text{inc}(t)=\text{e}^{j\omega_0{t}}$.}
     \label{FIG:Steady_Stage}
\end{figure}

This scenario is illustrated in \figref{FIG:Steady_Stage}, which shows that after the impulse response function has decayed to a sufficiently low level, about at time $T_\text{d}$, the output signal $\psi_\text{scat}(t)$ retrieves the waveform of the input harmonic signal $\psi_\text{inc}(t)=\text{e}^{j\omega_0{t}}$. Thus, after the transient time $T_\text{d}$, the system may be reasonably approximated as a time-invariant one, and the dispersion can be generally ignored at time scales larger than the modulation, $T_\text{b}\gg{T}_\text{m}$. In contrast, if the metasurface were switched at a time scale smaller than $T_\text{d}$, i.e., $T_\text{b}<{T}_\text{m}$, then the wave would see a change of reflection coefficient before reaching its steady state, and the system would really need to be described as a simultaneously time-variant \emph{and} dispersive one. In practice, the condition~\eqref{EQ:harmonicDemod} can be safely satisfied.

\section{Modulation Sequence}\label{Sec:Mod}
An ideal choice for the modulation of the proposed metasurface camouflaging system (\figref{FIG:Schematic}) would be an infinite-bandwidth white noise, since such a modulation, assuming finite energy, would lead to a uniform zero spectral density, and hence to perfect camouflaging. However, practically, the bandwidth of the modulation is limited by the speed of the switching elements, which will be PIN diodes in our experiment (Sec.~\ref{sec:exp_res}). Moreover, the selectivity functionality (Sec.~\ref{sec:selectivity}) and the interference immunity property (Sec.~\ref{sec:interf_imm}) of the system require some level of coherence, related to the condition~\eqref{EQ:condition}.

\begin{figure}[h!t]
    \centering
    \includegraphics{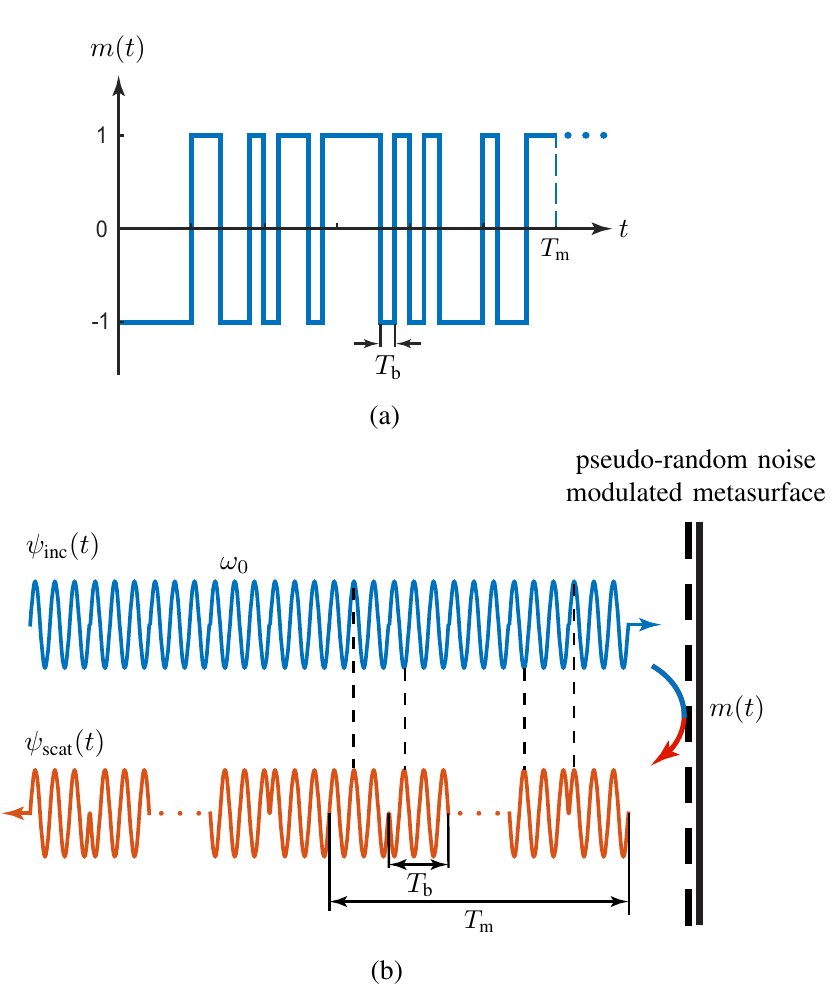}
     \caption{Practical modulation for the proposed system. (a)~ Modulation function, $m(t)$, consisting of a periodically repeated pseudo-random noise sequence of $N$ bits with bit duration $T_\text{b}$, and hence period $T_\text{m}=NT_\text{b}$ (one period shown). (b)~Corresponding scattered waveform.}
     \label{FIG:PN}
\end{figure}

We shall therefore use the pseudo-random noise periodic modulation scheme shown in \figref{FIG:PN} for $m(t)$. Figure~\ref{FIG:PN}(a) plots this  function. It consists of rectangular pulse pseudo-randomly oscillating between the values $+1$ and $-1$ at the bit rate or switching frequency of $f_\text{b}=1/T_\text{b}$, and with a bit period of $N$ bits, or time period of $T_\text{m}=NT_\text{b}$, corresponding to the function repetition frequency $f_\text{m}=1/T_\text{m}$. Figure~\ref{FIG:PN}(b) shows resulting scattered waveform, which is a harmonic wave with $\pi$-phase discontinuities corresponding to the switching times between the states $\pm{1}$.

The pseudo-random function, as the camouflaging key, should change from time to time to minimize the chances of foe radars to find it. Therefore, it does not have a uniquely determined spectrum, $\tilde{M}(\omega)$. However, the function $m(t)$ can be generally represented by its \emph{autocorrelation function}~\cite{Book:Torrieri_2018_SS}
\begin{subequations}\label{EQ:autocorr}
\begin{equation}
\begin{split}
s_\text{p}(t)
&=\int_{-\infty}^{+\infty}m(\tau)m(t+\tau)\,d\tau \\
&=-\frac{1}{N}+\frac{N+1}{N}\sum_{n=-\infty}^{+\infty}\Lambda\left(\frac{t-nNT_\text{b}}{T_\text{b}}\right),
\end{split}
\end{equation}
which is also of period $T_\text{m}=NT_\text{b}$, and where $\Lambda(\cdot)$ is the triangular function
\begin{equation}
\Lambda(t)=
\begin{cases}
1-|t|& \text{if}~t\leq 1,\\
0& \text{if}~t>1,
\end{cases}
\end{equation}
\end{subequations}
which essentially results from the correlation integral of the rectangular pulses composing $m(t)$ (\figref{FIG:PN}).

The Fourier transform of~\eqref{EQ:autocorr} is the \emph{power spectral density} function of $m(t)$, which reads~\cite{Book:Torrieri_2018_SS}
\begin{subequations}\label{EQ:PSD}
\begin{equation}
\tilde{s}_\text{p}(f)
=\frac{1}{N^2}\delta(f)+\frac{N+1}{N^2}\sum_{n=-\infty \atop n\neq0}^{+\infty}\text{sinc}^2\left(\frac{n}{N}\right)\delta\left(f-\frac{n}{N}f_\text{b}\right),
\end{equation}
and
\begin{equation}
\text{sinc}(f)=
\begin{cases}
0&  \text{if}~f=1,\\
\dfrac{\sin(\pi f)}{\pi f}& \text{otherwise}.
\end{cases}
\end{equation}
\end{subequations}

\begin{figure}[h!]
    \centering
    \includegraphics{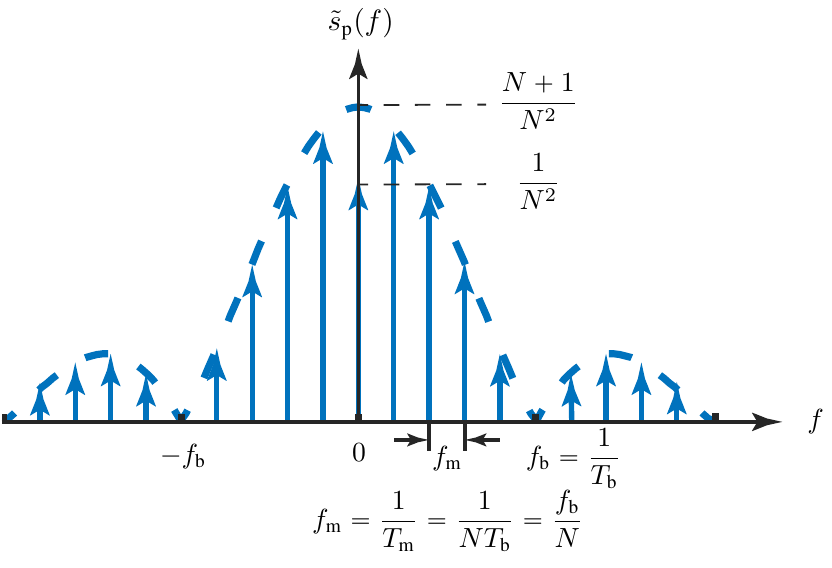}
     \caption{Power spectral density function $\tilde{s}_\text{p}(f)$, given by Eq.~\eqref{EQ:PSD}, for the modulation function $m(t)$ in \figref{FIG:PN}(a).}
     \label{FIG:PN_PSD}
\end{figure}

The power spectral density function~\eqref{EQ:PSD} $\tilde{s}_\text{p}(f)$ is plotted in \figref{FIG:PN_PSD}. This function is \emph{discrete}, due to the periodic nature of $s_\text{p}(t)$, with period $1/T_\text{m}=f_\text{m}=f_\text{b}/N=1/(NT_\text{b})$. It has the envelope $\frac{N+1}{N^2}\text{sinc}^2(f/f_\text{b})$, with maximum value $\frac{N+1}{N^2}$, main-lobe bandwidth $2f_\text{b}$, and DC value $1/N^2$.

This behavior may be explained as follows:
\begin{itemize}
	\item The nonzero $1/N^2$ value at $f=0$ (DC component) is due to the fact that $N$, as the \emph{length of a pseudo-random noise sequence}, is an odd number~\cite{Book:Torrieri_2018_SS}, so that there is always an imbalance between the number of $-1$ bits and the number of $+1$ bits, an effect that progressively vanishes by dilution as $N\rightarrow\infty$;
	\item As $N$ increases, assuming fixed switching frequency, the frequency interval between the function samples ($1/T_\text{m}=f_\text{m}=f_\text{b}/N$) reduces at the rate of $1/N$, and therefore the spectral sample density -- proportional to the number of samples within the first lobe of the sinc squared envelope -- increases at the same rate ($N$). If the input power is fixed, as may be assumed for given radar interrogator in the application of interest, then the power level of each sample must then be reduced by the same factor ($N$). This is indeed what is seen in \figref{FIG:PN_PSD}, considering that $\lim_{N\rightarrow\infty}(N+1)/N^2=1/N$. So, \emph{increasing the length of the pseudo-random noise sequence results in decreasing the level of the power spectrum density function}.
	\item Finally, decreasing the bit length ($T_\text{b}$), or equivalently increasing the bit rate ($f_\text{b}$), for a fixed sequence length ($N$), increases the fastest variation of $m(t)$ and hence \emph{spreads out its spectrum} -- in particular the spectral width of the sinc squared main lobe -- while decreasing the sample density, which depends only on the modulation period ($T_\text{m}$) at the same rate ($f_\text{b}$).
\end{itemize}

\section{Simulation Results}
Assuming an incident harmonic wave of frequency \mbox{$f_0=10$~GHz} (\figref{FIG:PN}), the power spectral density of the wave scattered by the modulated metasurface may be obtained from Eq.~\eqref{EQ:frequencydomain} as
\begin{equation}\label{EQ:PSD_scat}
\tilde{s}_\text{scat}(\omega)=|\tilde{\psi}_\text{scat}(\omega)|^2=|\tilde{\tilde{R}}(\omega-\omega_0,\omega_0)|^2=\tilde{s}_\text{p}(\omega-\omega_0),
\end{equation}
where $\tilde{s}_\text{p}(\omega)$ is given by~\eqref{EQ:PSD} with $f=\omega/(2\pi)$.

\begin{figure}[h!]
    \centering
    \includegraphics{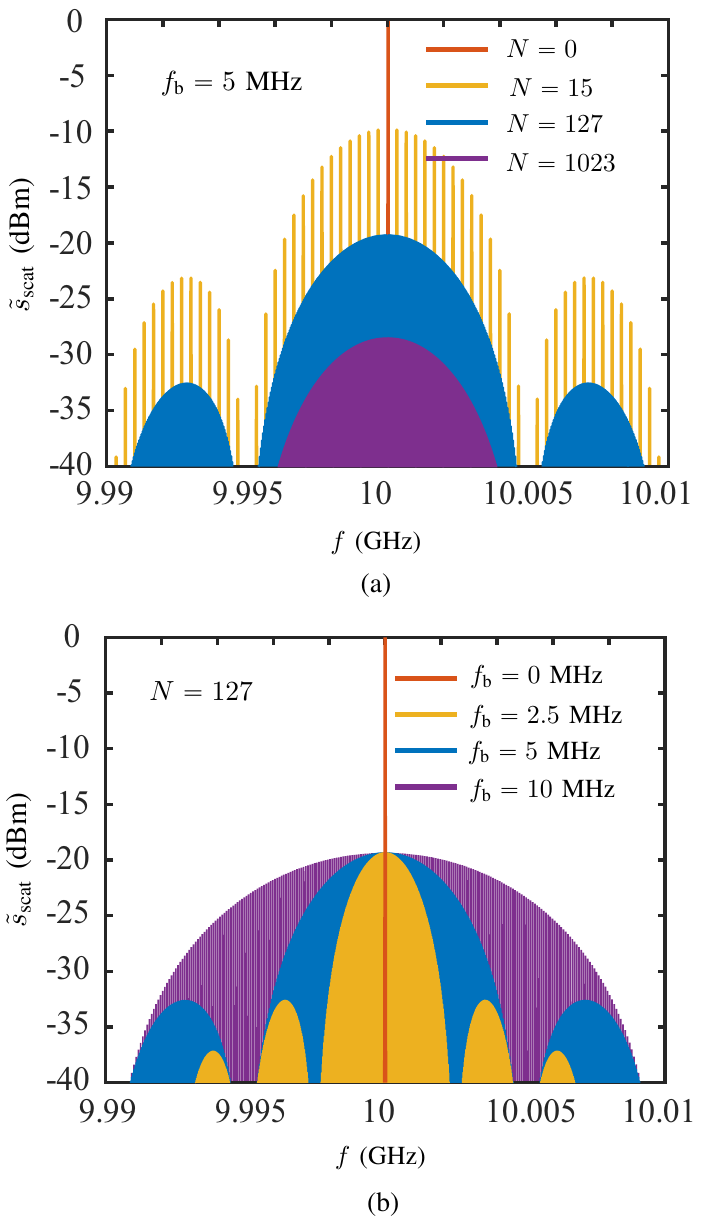}
     \caption{Parametric study of the power spectral density of a harmonic wave scattered by the metasurface, given by Eq.~\eqref{EQ:PSD_scat}. (a)~Spectrum level decrease with the increase of the modulation sequence length, $N$. (b)~Spectrum spreading with the increase of the modulation frequency, $f_\text{b}$.}
     \label{FIG:f_N}
\end{figure}

Figure~\ref{FIG:f_N} plots the power spectral density of the scattered wave for different values of the parameters $N$ and $f_\text{b}$ to illustrate the results of Sec.~\ref{Sec:Mod}. Figure~\ref{FIG:f_N}(a) shows how the level of the spectral power density decreases as $N$ increases, without change of bandwidth for fixed $f_\text{b}$. Figure~\ref{FIG:f_N}(b) shows how the bandwidth (main lobe) of the power spectral density increases as $f_\text{b}$ increases, without change of the maximum\footnote{Strictly, $\tilde{s}_\text{scat}(f_0)=1/N^2=6.2\cdot{10}^{-5}\approx{0}$ (or $-42.1$~dB, not visible in the figure), and the maximum refers here to the envelope maximum, which is very close to the level of the frequency samples nearest to $f_0$.} for fixed $N$.

\begin{figure}[h!]
    \centering
    \includegraphics{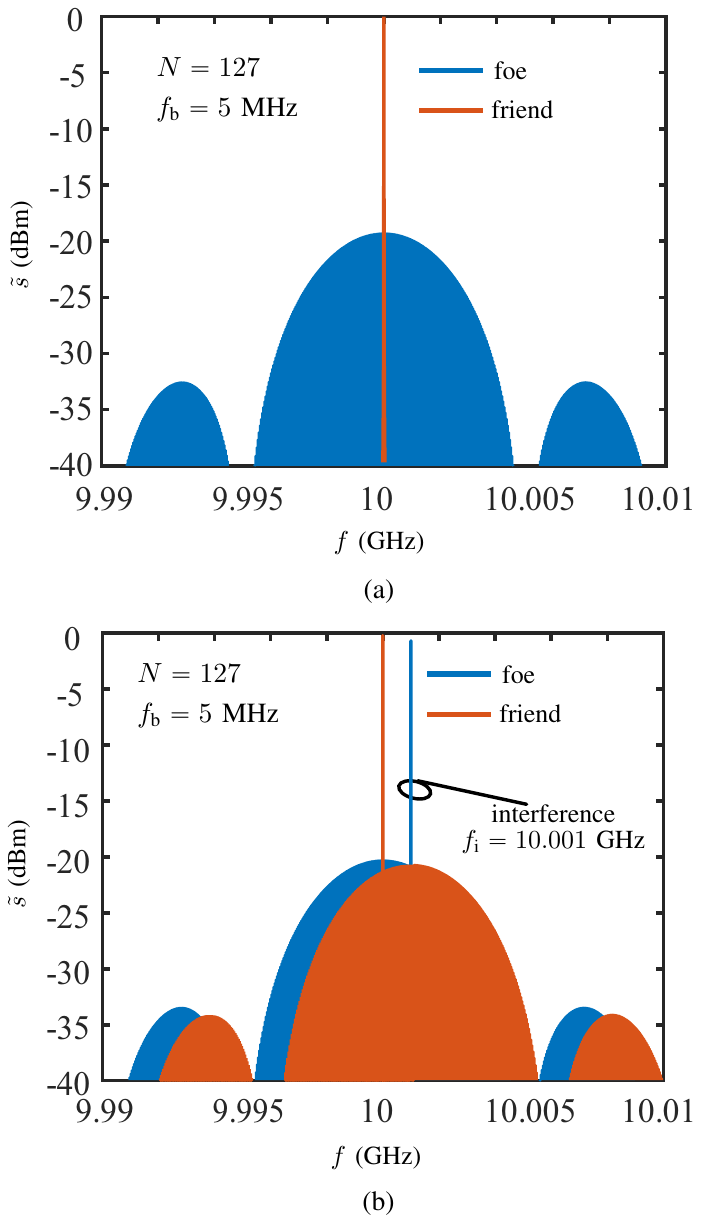}
     \caption{Comparison of the power spectral densities received by the foe radar and by the friend radar (\figref{FIG:SpreadSpectrum}).    	 (a)~Camouflaging selectivity, computed from Eq.~\eqref{EQ:harmonicDemod}. (b)~Interference immunity, computed from Eq.~\eqref{EQ:interfdemoduF}.}
     \label{FIG:demod}
\end{figure}

Figure~\ref{FIG:demod} compares the power spectral densities received by the foe radar and by the friend radar for the parameter pair $(N,f_\text{b})=(127,5~\text{MHz})$ (blue curves in \figref{FIG:f_N}). Figure~\ref{FIG:demod}(a) demonstrates the camouflaging selectivity of the metasurface system: the foe radar receives an undetectable spectrum spread signal, whereas the friend radar perfectly detects the object covered by the metasurface. Figure~\ref{FIG:demod}(b) demonstrates the interference immunity of the metasurface system: the foe radar detects only the interference signal, a harmonic wave at $10.001$~GHz, which appropriately delivers false information about the object, whereas the friend radar does not see the interference signal while still perfectly detecting the object.

\section{Experiment Results}\label{sec:exp_res}
\begin{figure}[h!]
    \centering
    \includegraphics{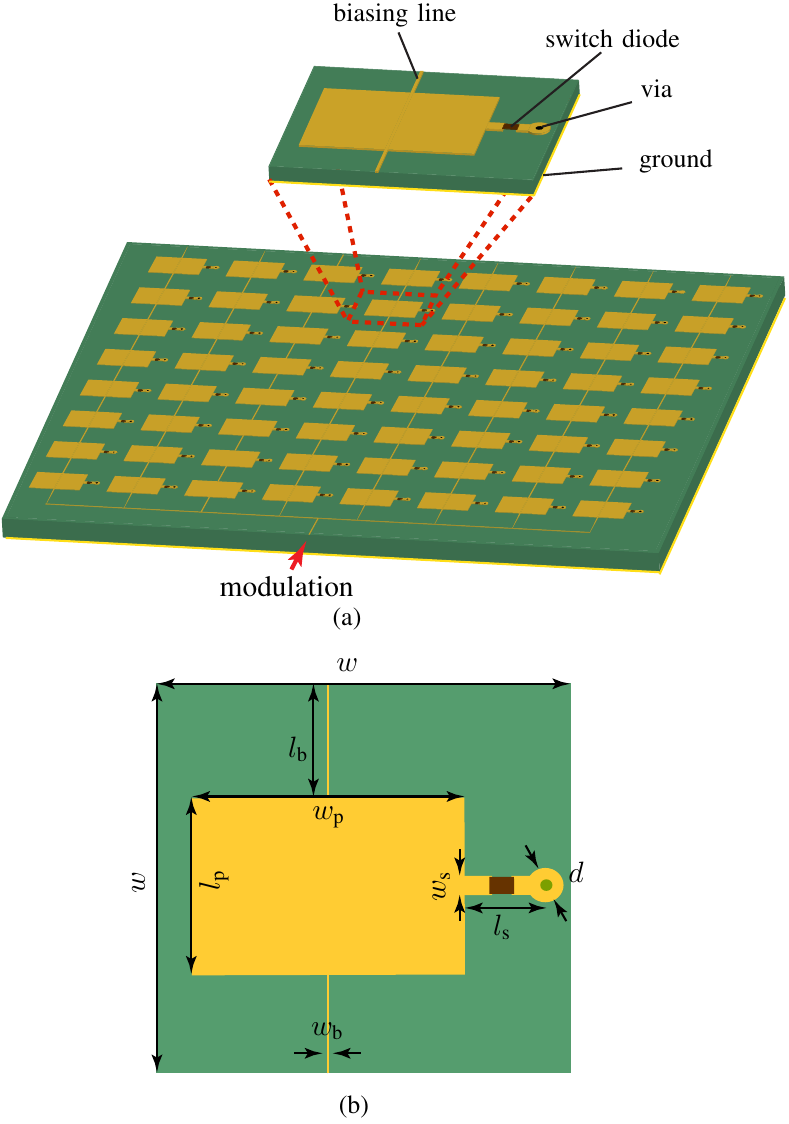}
     \caption{Layout of the metasurface prototype. (a)~Overall view. (b)~Unit cell. The parameter values are $w =15$~mm, \mbox{$w_\text{p}=7.6$~mm} (resonant length), $l_\text{p}=5.6$~mm, $l_\text{b}=4.7$~mm, $w_\text{s}=0.5$~mm, $w_\text{b}=0.2$~mm, $l_\text{s}=1.8$~mm and $d=0.4$~mm.}
     \label{FIG:Layout}
\end{figure}

Figure~\ref{FIG:Layout} shows the layout of the metasurface prototype. The metasurface is designed on a Rogers~6002 substrate with permittivity of 2.94 and thickness 0.76~mm. Figure~\ref{FIG:Layout}(a) shows the overall metasurface. The bottom side of it is a ground plane and the top side is an $8\times8$ array of scatterers interconnected by bias lines that are meet to a single modulation point at one side of the structure. The unit cell consists of a rectangular patch, with the bias line connection at both sides of it and a PIN diode switch (MACOM MADP-00090714020) interconnecting the patch and the ground plane through a shorted metalized via. Figure~\ref{FIG:Layout}(b) provides the parameter values of the unit cell for operation at $10$~GHz.

\begin{figure}[h!]
    \centering
    \includegraphics{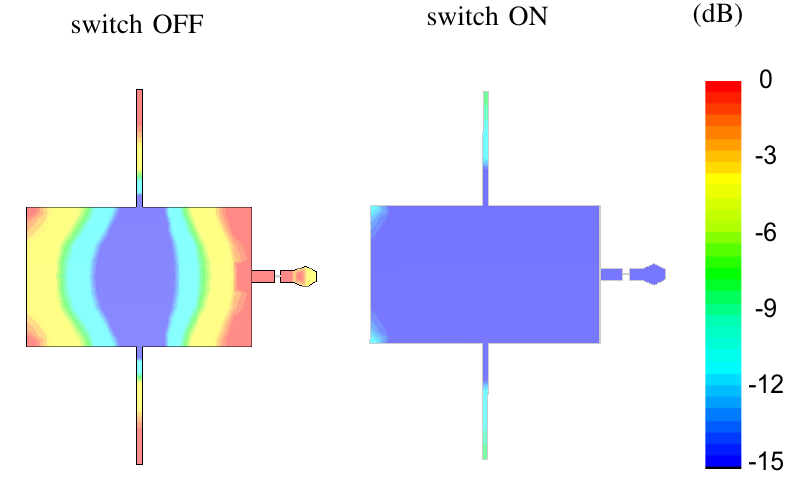}
     \caption{Simulated (FDTD -- CST Microwave Studio) surface current distribution on the patch of the unit cell in \figref{FIG:Layout} for the PIN diodes switched to the ON and OFF states.}
     \label{FIG:Field}
\end{figure}

Figure~\ref{FIG:Field} shows the simulated surface current distribution on the unit-cell patch for the PIN diodes switched to the ON and OFF states. When the diodes are OFF, the patches are isolated from the ground and resonate, which provides quasi-PMC reflection, while  when the diodes are ON, the patches are shorted to the ground, which provides a quasi-PEC reflection.

\begin{figure}[h!]
	\centering
    \includegraphics{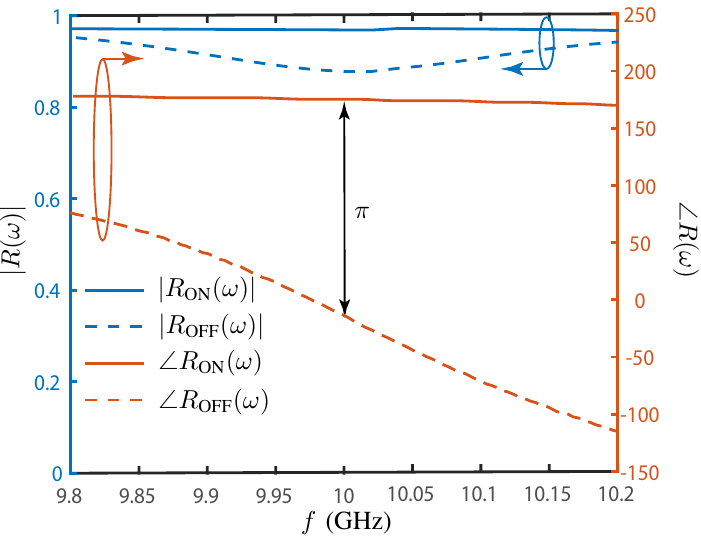}
	\caption{Reflection coefficient of the metasurface prototoype in \figref{FIG:Layout} for the PIN diodes switched to the ON and OFF states, corresponding to the current distributions in \figref{FIG:Field}.
	}
	\label{FIG:SP}
\end{figure}

Figure~\ref{FIG:SP} plots the reflection coefficients for the 2 modulation states of the metasurface. The amplitude of this coefficient is close to 1 for both states, while the phase difference between the two states is around $\pi$ at the operation frequency (10~GHz), as desired. The off state features a slightly larger loss than on state, due to its resonant nature, which is apparent in~\figref{FIG:Field}. The slight loss in the on state is mostly due to the resistive loss in the diodes.

The imbalance between the on- and off-state reflective amplitudes (\figref{FIG:SP}) may affect the camouflaging performance. Indeed, this imperfection typically alters the balance between the $+1$ (on-state) and $-1$ bits (off-state) in the modulation function [\figref{FIG:PN}(a)]. This introduces a DC component in the scattered wave (here positive due to the higher reduction of the $-1$ compared to the $+1$ bits), and hence at the center frequency of the metasurface, which could ultimately reveal to the foe the presence of the object. The level of imbalance may be reduced by adding a resistor in series with PIN diode or by using a substrate of lower loss. However, the level of unbalanced seen in \figref{FIG:Field} was found to be acceptable for the current proof of concept, as will be seen shortly.

\begin{figure}[h!]
    \centering
    \includegraphics{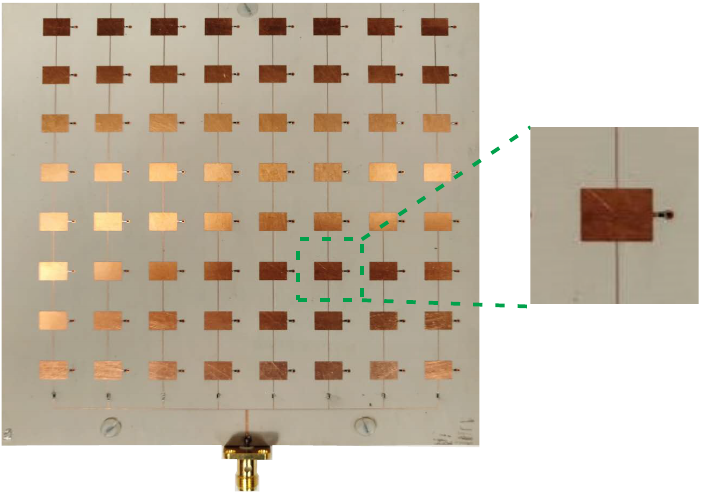}
     \caption{Fabricated metasurface prototype, corresponding to \figref{FIG:Layout}.}
     \label{FIG:Prototype}
\end{figure}

Figure~\ref{FIG:Prototype} shows the fabricated 10-GHz prototype while \figref{FIG:SP} shows the experimental set-up used to demonstrate the metasurface system. The metasurface is placed on a piece of absorbing material on the floor. It is modulated by a pseudo-random noise sequence of length $N=127$ and rate $f_\text{b}=5$~MHz provided by an arbitrary signal generator. A pair of planar Vivaldi antennas, placed above the metasurface in the far-field, simulates an arbitrary interrogating radar. The 10~GHz harmonic wave is generated by a signal generator and sent towards the metasurface by the transmitting antenna. The wave scattered by the metasurface is then collected by the receiving antenna, passed through a bandpass filter that suppresses the out-of-band noise, and measured by a vector signal analyzer.
\begin{figure}[h!]
    \centering
    \includegraphics{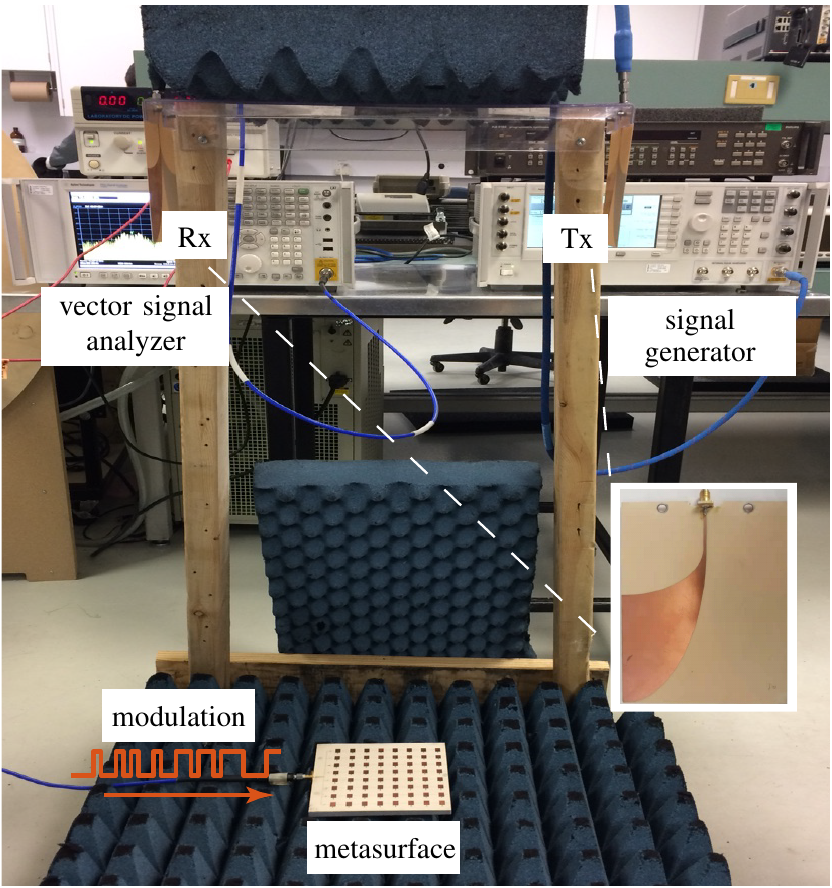}
     \caption{Experimental set-up.}
     \label{FIG:ExpSet}
\end{figure}

\begin{figure}[h!]
    \centering
    \includegraphics{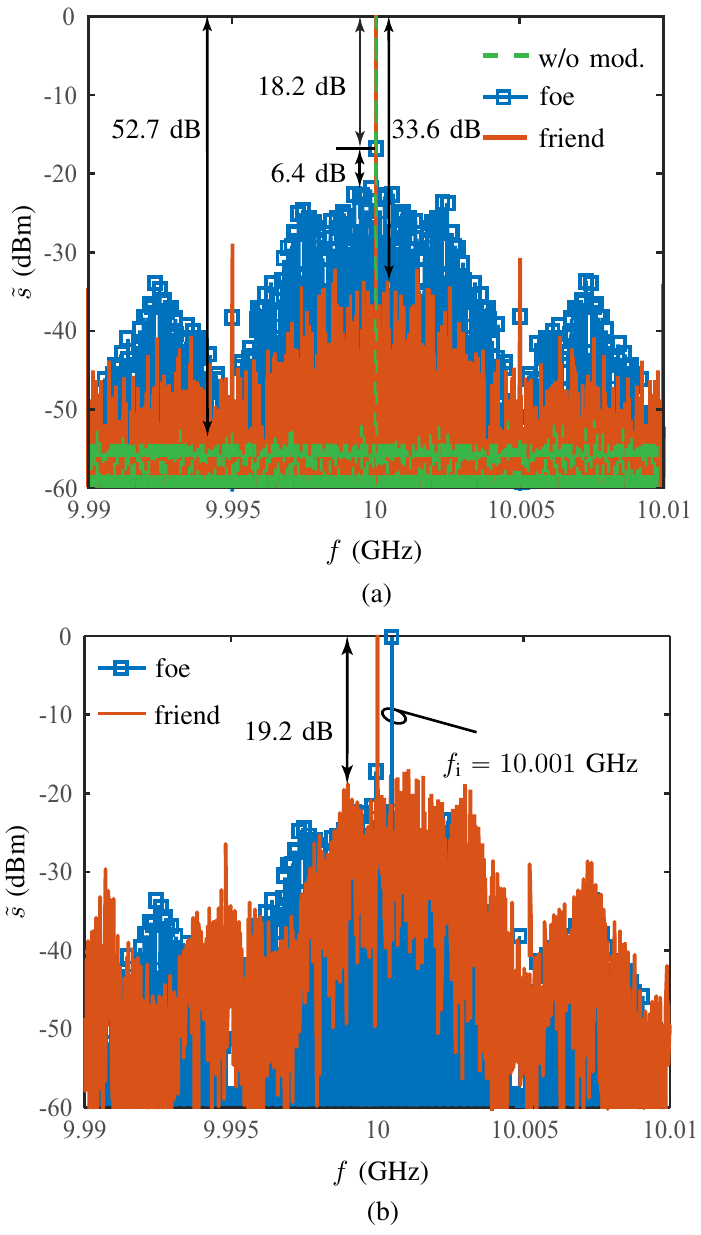}
     \caption{ Experimental results, corresponding to the simulations in \figref{FIG:demod}. (a)~Camouflaging selectivity. (b)~Interference immunity.}
     \label{FIG:ExpRes}
\end{figure}

Figure~\ref{FIG:ExpRes} plots the power spectral densities of the signals detected by the foe and friend radars. Figure~\ref{FIG:ExpRes}(a) demonstrates the camouflaging selectivity of the system. When the metasurface is not modulated, the receiver perfectly detects the 10~GHz harmonic wave sent by the transmitter and scattered by the metasurface (green curve), with a signal-to-noise ratio (SNR) of 52.7~dB. When the metasurface is modulated, the scattered wave is spread out into a relatively broad band signal of 10~MHz bandwidth and with level reduced by 18.2~dB by the spectrum spreading operation. The scattered signal is not perfectly removed here; its 6.4~dB SNR level is due to several factors, including spurious coupling between transmitting and receiving antennas, scattering from the objects in the room, and, to a lesser extend, the imperfect amplitude balance between the two states of the metasurface~\figref{FIG:SP}. However, this experiment clearly demonstrates the camouflaging proof-of-concept of the metasurface system. Upon demodulation, the friend radar recovers the transmitted harmonic signal, with an SNR of 33.6~dB,
and hence detects the object, which demonstrates the selectivity property of the system.

Figure~\ref{FIG:ExpRes}(b) shows the interference response of the system to an interfering signal of $f_\text{i}=10.001$~GHz. Whereas the foe radar is still unable to detect the object and additionally strongly detects the interfering signal, the friend radar does not detect this signal at all while still detecting the object, with an SNR of around 19.2~dB.

\section{Conclusion}
A time-modulated metasurface-based camouflaging technology has been proposed, analyzed and experimentally demonstrated. Given its spread spectrum, selectivity and interference immunity features, as well as its potential efficiency, this technology may find wide applications in both defense and civilian applications.

\bibliographystyle{IEEEtran}
\bibliography{IEEEabrv,Xiaoyi_ST_Reference}

\begin{thebibliography}{10}
\providecommand{\url}[1]{#1}
\csname url@samestyle\endcsname
\providecommand{\newblock}{\relax}
\providecommand{\bibinfo}[2]{#2}
\providecommand{\BIBentrySTDinterwordspacing}{\spaceskip=0pt\relax}
\providecommand{\BIBentryALTinterwordstretchfactor}{4}
\providecommand{\BIBentryALTinterwordspacing}{\spaceskip=\fontdimen2\font plus
\BIBentryALTinterwordstretchfactor\fontdimen3\font minus
  \fontdimen4\font\relax}
\providecommand{\BIBforeignlanguage}[2]{{%
\expandafter\ifx\csname l@#1\endcsname\relax
\typeout{** WARNING: IEEEtran.bst: No hyphenation pattern has been}%
\typeout{** loaded for the language `#1'. Using the pattern for}%
\typeout{** the default language instead.}%
\else
\language=\csname l@#1\endcsname
\fi
#2}}
\providecommand{\BIBdecl}{\relax}
\BIBdecl

\bibitem{Book:Stevens_2011_AnimalCamouflage}
M.~Stevens and S.~Merilaita, \emph{Animal Camouflage: Mechanisms and
  Function}.\hskip 1em plus 0.5em minus 0.4em\relax Cambridge University Press,
  2011.

\bibitem{Book:Lynch_RF_Stealth}
D.~D. Lynch and I.~of~Electrical~Engineers, \emph{Introduction to RF
  Stealth}.\hskip 1em plus 0.5em minus 0.4em\relax SciTech, 2004.

\bibitem{Jour:Rune_2018_VisualCamouflage}
R.~Pettersson, ``Visual camouflage,'' \emph{J. Vis. Lit.}, vol.~37, no.~3, pp.
  181--194, 2018.

\bibitem{Jour:APL_Wang_2011_AbsorbingMaterial}
C.~Wang, X.~Han, P.~Xu, X.~Zhang, Y.~Du, S.~Hu, J.~Wang, and X.~Wang, ``The
  electromagnetic property of chemically reduced graphene oxide and its
  application as microwave absorbing material,'' \emph{Appl. Phys. Lett.},
  vol.~98, no.~7, p. 072906, 2011.

\bibitem{Jour:AM_Xia_2013_AbsorbingMaterial}
T.~Xia, C.~Zhang, N.~A. Oyler, and X.~Chen, ``Hydrogenated tio$_2$
  nanocrystals: A novel microwave absorbing material,'' \emph{Adv. Mater.},
  vol.~25, no.~47, pp. 6905--6910, 2013.

\bibitem{Jour:JMMM_Lima_2008_RadarAbsorbingMaterial}
U.~Lima, M.~Nasar, R.~Nasar, M.~Rezende, and J.~Araújo, ``Ni–zn nanoferrite
  for radar-absorbing material,'' \emph{J. Magn. Magn. Mater.}, vol. 320,
  no.~10, pp. 1666 -- 1670, 2008.

\bibitem{Jour:TM_Bongeson_2004_RCSOptimaization}
A.~{Bondeson}, Y.~{Yang}, and P.~{Weinerfelt}, ``Optimization of radar cross
  section by a gradient method,'' \emph{IEEE Trans. Magn}, vol.~40, no.~2, pp.
  1260--1263, March 2004.

\bibitem{Jour:Tennant_2004_Screen}
B.~Chambers and A.~Tennant, ``The phase-switched screen,'' \emph{IEEE Antenn.
  Prooag. M.}, vol.~46, no.~6, pp. 23--37, 2004.

\bibitem{Jour:Tennant_1997_Screen}
A.~Tennant, ``Reflection properties of a phase modulating planar screen,''
  \emph{Electron. Lett.}, vol.~33, no.~21, pp. 1768--1769, 1997.

\bibitem{Jour:Nanophotonics_2018_Achouri_Metasurface}
K.~Achouri and C.~Caloz, ``Design, concepts, and applications of
  electromagnetic metasurfaces,'' \emph{Nanophotonics}, vol.~7, no.~6, pp.
  1095--1116, 2018.

\bibitem{Jour:PRX_jiang_2014_ControllingPolarization}
S.-C. Jiang, X.~Xiong, Y.-S. Hu, Y.-H. Hu, G.-B. Ma, R.-W. Peng, C.~Sun, and
  M.~Wang, ``Controlling the polarization state of light with a dispersion-free
  metastructure,'' \emph{Phys. Rev. X}, vol.~4, no.~2, p. 021026, 2014.

\bibitem{Jour:PJ_Genevet_2014_WavefrontControl}
P.~Genevet and F.~Capasso, ``Breakthroughs in photonics 2013: Flat optics:
  Wavefronts control with huygens' interfaces,'' \emph{IEEE Photon. Technol.
  Lett.}, vol.~6, no.~2, pp. 1--4, 2014.

\bibitem{Jour:PRL_Asadchy_2015_FunctionalMetamirrors}
V.~S. Asadchy, Y.~Ra’Di, J.~Vehmas, and S.~Tretyakov, ``Functional
  metamirrors using bianisotropic elements,'' \emph{Phys. Rev. Lett.}, vol.
  114, no.~9, p. 095503, 2015.

\bibitem{Jour:LSA_Karimi_2014_OAM}
E.~Karimi, S.~A. Schulz, I.~De~Leon, H.~Qassim, J.~Upham, and R.~W. Boyd,
  ``Generating optical orbital angular momentum at visible wavelengths using a
  plasmonic metasurface,'' \emph{Light: Science \& Applications}, vol.~3,
  no.~5, p. e167, 2014.

\bibitem{Jour:PRApp_Pfeiffer_2014_BesselBeam}
C.~Pfeiffer and A.~Grbic, ``Controlling vector bessel beams with
  metasurfaces,'' \emph{Phys. Rev. Appl.}, vol.~2, no.~4, p. 044012, 2014.

\bibitem{Jour:NN_Zheng_2015_MetasurfaceHolograms}
G.~Zheng, H.~M{\"u}hlenbernd, M.~Kenney, G.~Li, T.~Zentgraf, and S.~Zhang,
  ``Metasurface holograms reaching 80\% efficiency,'' \emph{Nat. Nanotechnol.},
  vol.~10, no.~4, p. 308, 2015.

\bibitem{Jour:NC_YE_2016_NonlinearMetasurfaceHolography}
W.~Ye, F.~Zeuner, X.~Li, B.~Reineke, S.~He, C.-W. Qiu, J.~Liu, Y.~Wang,
  S.~Zhang, and T.~Zentgraf, ``Spin and wavelength multiplexed nonlinear
  metasurface holography,'' \emph{Nat. Commun.}, vol.~7, p. 11930, 2016.

\bibitem{Jour:Sounas_2012_NonreciprocalMetasurface}
D.~L. Sounas, T.~Kodera, and C.~Caloz, ``Electromagnetic modeling of a
  magnetless nonreciprocal gyrotropic metasurface,'' \emph{IEEE Trans. Antennas
  Propag.}, vol.~61, no.~1, pp. 221--231, 2012.

\bibitem{Jour:Caloz_2018_PRApp_Nonreciprocity}
C.~Caloz, A.~Al\`u, S.~Tretyakov, D.~Sounas, K.~Achouri, and Z.-L.
  Deck-L\'eger, ``Electromagnetic nonreciprocity,'' \emph{Phys. Rev. Applied},
  vol.~10, p. 047001, Oct 2018.

\bibitem{Jour:Grbic_2015_PRB_tractorbeam}
C.~Pfeiffer and A.~Grbic, ``Generating stable tractor beams with dielectric
  metasurfaces,'' \emph{Phys. Rev. B}, vol.~91, p. 115408, Mar 2015.

\bibitem{Jour:Achouri_2019_TAP_OpticalForce}
K.~{Achouri}, O.~V. {Cespedes}, and C.~{Caloz}, ``Solar “meta-sails” for
  agile optical force control,'' \emph{IEEE Trans. Antennas Propag.}, pp. 1--1,
  2019.

\bibitem{Jour:Science_Silva_2014_MathOperation}
A.~Silva, F.~Monticone, G.~Castaldi, V.~Galdi, A.~Al{\`u}, and N.~Engheta,
  ``Performing mathematical operations with metamaterials,'' \emph{Science},
  vol. 343, no. 6167, pp. 160--163, 2014.

\bibitem{Jour:TAP_1973_Emerson_EM_absorbers}
W.~Emerson, ``Electromagnetic wave absorbers and anechoic chambers through the
  years,'' \emph{IEEE Trans. Antennas Propag.}, vol.~21, no.~4, pp. 484--490,
  1973.

\bibitem{Jour:JPDAP_li_2014_CircAbsorber}
M.~Li, L.~Guo, J.~Dong, and H.~Yang, ``An ultra-thin chiral metamaterial
  absorber with high selectivity for {L}{C}{P} and {R}{C}{P} waves,'' \emph{J.
  Phys. D}, vol.~47, no.~18, p. 185102, 2014.

\bibitem{Jour:PRB_Valagiannopoulos_2015_PerfectMirrorAndAborber}
C.~Valagiannopoulos, A.~Tukiainen, T.~Aho, T.~Niemi, M.~Guina, S.~Tretyakov,
  and C.~Simovski, ``Perfect magnetic mirror and simple perfect absorber in the
  visible spectrum,'' \emph{Phys. Rev. B}, vol.~91, no.~11, p. 115305, 2015.

\bibitem{Jour:OL_Wen_2014_AbsorptionMetamaterial}
Y.~Wen, W.~Ma, J.~Bailey, G.~Matmon, X.~Yu, and G.~Aeppli, ``Planar broadband
  and high absorption metamaterial using single nested resonator at terahertz
  frequencies,'' \emph{Opt. Lett.}, vol.~39, no.~6, pp. 1589--1592, 2014.

\bibitem{Jour:EM_1985_Raj_Scattering}
T.~Cwik and R.~Mittra, ``Scattering from general periodic screens,''
  \emph{Electromagnetics}, vol.~5, no.~4, pp. 263--283, 1985.

\bibitem{Jour:SR_2016_Chen_Redirection_metasurface}
K.~Chen, Y.~Feng, Z.~Yang, L.~Cui, J.~Zhao, B.~Zhu, and T.~Jiang, ``Geometric
  phase coded metasurface: from polarization dependent directive
  electromagnetic wave scattering to diffusion-like scattering,'' \emph{Sci.
  Rep.}, vol.~6, p. 35968, 2016.

\bibitem{Jour:PRB_2009_Alu_MantleCloak}
A.~Al\`{u}, ``Mantle cloak: {I}nvisibility induced by a surface,'' \emph{Phys.
  Rev. B}, vol.~80, p. 245115, Dec 2009.

\bibitem{Jour:TAP_Dehmollaian_2019_Cloaking}
M.~{Dehmollaian}, N.~{Chamanara}, and C.~{Caloz}, ``Wave scattering by a
  cylindrical metasurface cavity of arbitrary cross section: Theory and
  applications,'' \emph{IEEE Trans. Antennas Propag.}, vol.~67, no.~6, pp.
  4059--4072, June 2019.

\bibitem{Jour:Alu_AWPL_2014_CarpetCloak}
N.~M. {Estakhri} and A.~{Alù}, ``Ultra-thin unidirectional carpet cloak and
  wavefront reconstruction with graded metasurfaces,'' \emph{IEEE Antennas
  Wirel. Propag. Lett}, vol.~13, pp. 1775--1778, 2014.

\bibitem{Jour:Fleury_2015_invisibility}
R.~Fleury, F.~Monticone, and A.~Al{\`u}, ``Invisibility and cloaking: Origins,
  present, and future perspectives,'' \emph{Phys. Rev. Appl.}, vol.~4, no.~3,
  p. 037001, 2015.

\bibitem{Jour:ArXiv_Caloz_2019_SpacetimeMeta}
C.~Caloz and Z.-L. Deck-L{\'e}ger, ``Spacetime metamaterials,'' \emph{arXiv
  preprint arXiv:1905.00560}, 2019.

\bibitem{Jour:Arxiv_Wu_2019_frequency_translation}
Z.~Wu and A.~Grbic, ``Serrodyne frequency translation using time-modulated
  metasurfaces,'' \emph{arXiv preprint arXiv:1905.06792}, 2019.

\bibitem{Jour:CC_Cui_2019_CommMeta}
W.~{Tang}, X.~{Li}, J.~Y. {Dai}, S.~{Jin}, Y.~{Zeng}, Q.~{Cheng}, and T.~J.
  {Cui}, ``Wireless communications with programmable metasurface: Transceiver
  design and experimental results,'' \emph{China Commun.}, vol.~16, no.~5, pp.
  46--61, May 2019.

\bibitem{Jour:AMT_Cui_2019_CommMeta}
J.~Y. Dai, W.~K. Tang, J.~Zhao, X.~Li, Q.~Cheng, J.~C. Ke, M.~Z. Chen, S.~Jin,
  and T.~J. Cui, ``Metasurfaces: Wireless communications through a simplified
  architecture based on time-domain digital coding metasurface,'' \emph{Adv.
  Mater. Technol.}, vol.~4, no.~7, p. 1970037, 2019.

\bibitem{Jour:Res_Cui_2019_PrograCoding}
T.~J. Cui, S.~Liu, G.~D. Bai, Q.~Ma \emph{et~al.}, ``Direct transmission of
  digital message via programmable coding metasurface,'' \emph{Research}, p.
  2584509, 2019.

\bibitem{Jour:Cui_NSR_2018_Comm}
J.~Zhao, X.~Yang, J.~Y. Dai, Q.~Cheng, X.~Li, N.~H. Qi, J.~C. Ke, G.~D. Bai,
  S.~Liu, S.~Jin, A.~Alù, and T.~J. Cui, ``Programmable time-domain
  digital-coding metasurface for non-linear harmonic manipulation and new
  wireless communication systems,'' \emph{Natl. Sci. Rev.}, vol.~6, no.~2, pp.
  231--238, 11 2018.

\bibitem{Conf:Wang_2019_APS_DOA}
X.~{Wang} and C.~{Caloz}, ``Direction-of-arrival (doa) estimation based on
  spacetime-modulated metasurface,'' in \emph{IEEE AP-S Int. Antennas Propag.
  (APS)}, Atlanta, USA, July 2019.

\bibitem{Jour:OME_2015_Shalaev_Nonreciprocity}
A.~Shaltout, A.~Kildishev, and V.~Shalaev, ``Time-varying metasurfaces and
  {L}orentz non-reciprocity,'' \emph{Opt. Mater. Express}, vol.~5, no.~11, pp.
  2459--2467, 2015.

\bibitem{Jour:PRX_2018_shadrivov_STM}
M.~Liu, D.~A. Powell, Y.~Zarate, and I.~V. Shadrivov, ``Huygens’ metadevices
  for parametric waves,'' \emph{Phys. Rev. X}, vol.~8, no.~3, p. 031077, 2018.

\bibitem{Jour:PRB_2017_Ramaccia_Doppler_cloak}
D.~Ramaccia, D.~L. Sounas, A.~Al\`u, A.~Toscano, and F.~Bilotti, ``Doppler
  cloak restores invisibility to objects in relativistic motion,'' \emph{Phys.
  Rev. B}, vol.~95, p. 075113, Feb 2017.

\bibitem{Jour:TAP_2019_Nima_STM}
N.~Chamanara, Y.~Vahabzadeh, and C.~Caloz, ``Simultaneous control of the
  spatial and temporal spectra of light with space-time varying metasurfaces,''
  \emph{IEEE Trans. Antennas Propag.}, vol.~67, no.~4, pp. 2430--2441, April
  2019.

\bibitem{Conf:Wang_2019_APS_Camouflaging}
X.~{Wang} and C.~{Caloz}, ``Spread-spectrum camouflaging based on
  time-modulated metasurface,'' in \emph{IEEE AP-S Int. Antennas Propag.
  (APS)}, Atlanta, USA, July 2019.

\bibitem{Jour:Lathi_2005_LinearSysSignals}
B.~P. Lathi and R.~A. Green, \emph{Linear Systems and Signals}.\hskip 1em plus
  0.5em minus 0.4em\relax Oxford University Press New York, 2005.

\bibitem{Book:Torrieri_2018_SS}
D.~Torrieri, \emph{Principles of Spread-Spectrum Communication Systems}.\hskip
  1em plus 0.5em minus 0.4em\relax Springer, 2018.

\end{thebibliography}

\end{document}